\documentclass[12pt,preprint]{aastex}

\begin{document}
 
\title{
A BeppoSAX Observation of KS 1731--260 in its Quiescent State:
Constraints on the NS Magnetic Field
}

\author{L. Burderi\altaffilmark{1}, T. Di Salvo\altaffilmark{2,3}, 
L. Stella\altaffilmark{1}, F. Fiore\altaffilmark{1},
N.R. Robba\altaffilmark{3}, 
M. van der Klis\altaffilmark{2}, R. Iaria\altaffilmark{3}, 
M. Mendez\altaffilmark{4}, M. T. Menna\altaffilmark{1}, 
S. Campana\altaffilmark{5}, 
G. Gennaro\altaffilmark{6}, S. Rebecchi\altaffilmark{7},
M. Burgay\altaffilmark{8}}
\altaffiltext{1}{Osservatorio Astronomico di Roma, Via Frascati 33, 
00040 Monteporzio Catone (Roma), Italy; burderi@coma.mporzio.astro.it.}
\altaffiltext{2}{Astronomical Institute "Anton Pannekoek," University of 
Amsterdam and Center for High-Energy Astrophysics,
Kruislaan 403, NL 1098 SJ Amsterdam, the Netherlands; disalvo@astro.uva.nl.}
\altaffiltext{3}{Dipartimento di Scienze Fisiche ed Astronomiche, 
Universit\`a di Palermo, Via Archirafi 36, 90123 Palermo, Italy.}
\altaffiltext{4}{SRON, Laboratory for Space Research, Sorbonnelaan 2, 
NL-3584 CA, Utrecht, The Netherlands.}
\altaffiltext{5}{Osservatorio Astronomico di Brera, Via Bianchi 46, I-23807 
Merate, Italy.}
\altaffiltext{6}{"Telespazio" S.p.A.}
\altaffiltext{7}{C.I.F.S./ASI Consorzio Interuniversitario per la Fisica
Spaziale/Agenzia Spaziale Italiana.}
\altaffiltext{8}{Dipartimento di Astronomia, Universit\`a di Bologna, 
via Ranzani 1, 40127 Bologna, Italy.}

\begin{abstract}

We report here the results of a 90 ks BeppoSAX observation 
of the low mass X-ray binary and atoll source KS 1731--260 during a 
quiescent phase. From this observation we derive a 
source X-ray luminosity of $\sim 10^{33}$ ergs/s (for a 
source distance of 7 kpc).  If the neutron star is
spinning at a period of a few milliseconds, as inferred from the 
nearly-coherent oscillations observed during type-I X-ray bursts, the 
quiescent X-ray luminosity constrains the neutron star
magnetic field strength.  We consider all the mechanisms that have been 
proposed to explain the quiescent X-ray emission of neutron star X-ray
transients and compare the corresponding expectations with the measured 
upper limit on the X-ray luminosity. We find that, in any case, the neutron 
star magnetic field is most probably less than $\sim 10^{9}$ Gauss. 
We have also observed KS 1731--260, still in its quiescence state, at 1.4 GHz 
with the Parkes radiotelescope to search for radio pulses. 
We found that no radio signals with millisecond periods are present with 
an upper limit on the flux of 0.60 mJy using a 4 min integration 
time (optimal for a close system with an orbital period smaller than a few
hours) and of 0.21 mJy using a 35 min integration time (optimal for a wide 
orbit system).

\end{abstract}


\section{Introduction}

The Galactic X--ray source KS~1731--260 was discovered in 
August 1989 with the imaging spectrometer TTM on the Mir--Kvant observatory 
(Sunyaev 1989; Sunyaev et al. 1990).  During the $\sim 15$~days long 
Mir--Kvant observation the source intensity varied from 50 to 100 mCrab 
in the 2--27~keV band. The presence of three type--I
bursts indicated that the compact object is an accreting neutron
star (hereafter NS) and that the source distance is about 7 kpc
(Muno et al. 2000).
The factor of 10 intensity variations displayed by the 
source suggested that  KS~1731--260 is a transient source (e.g.\
Sunyaev et al. 1990). There were numerous detections of the source 
at a level of 50--100 mCrab since its discovery in 1989, and the monitoring 
by the Rossi X-ray Timing Explorer (RXTE) All Sky Monitor (ASM) showed that 
the source was continuously active 
until February 2001. Therefore the source appeared to belong to 
a group of NS soft X-ray transients (hereafter SXTs) that display active 
states extending for a number of years (as opposed to outbursts decaying 
on timescales of a few months; see Campana et al. 1998b for a review). 

The source spectrum resembles that of other SXTs in outburst; 
the TTM spectrum taken on August 16--31, 1989 
could be fit to a thermal bremsstrahlung with $kT \sim 5.7$ keV 
while the spectrum obtained by SIGMA on  March 14, 1991 
(when the source was somewhat fainter, $9 \times 10^{36}$ erg/s 
as opposed to $1.5 \times 10^{37}$ erg/s) was well modelled by a 
power law with a photon index of $\sim 2.9$ extending up to 150 keV at 
least.  
The source detection during the ROSAT all-sky survey yielded a fairly accurate 
measurement of column density, $N_H = (1.00 \pm 0.19) \times 10^{22}$ cm$^{-2}$
(Barret, Motch, \& Predehl 1998).

Based on an RXTE/PCA observation in July 1996,  
Smith, Morgan, \& Bradt (1997) discovered a nearly coherent 
signal at $523.92 \pm 0.05$ Hz (corresponding to a period of 1.91 ms) 
during a two-second interval close to the beginning of the decay of 
a type I X-ray burst.
This signal, as well as similar signals observed during type I X-ray bursts 
from about ten low mass X-ray binaries, likely corresponds to the NS 
rotation frequency (or twice its value; for a review see Strohmayer, Swank, 
\& Zhang 1998). 
An RXTE observation in August 1, 1996 (Wijnands \& van der Klis 1997) 
revealed two simultaneous quasi periodic oscillations (QPO) in the persistent 
emission of KS 1731--260 
at 898 Hz and 1159 Hz, the frequency separation of which ($260.3 \pm 9.6$ Hz) 
is compatible with half the frequency of the nearly-coherent signal 
observed in the type--I bursts. In the beat
frequency model for the kHz QPOs, this corresponds to a NS spinning
at a period of $3.8$ ms, which might imply that only the first harmonic is 
detected during type--I bursts (although this case seems to be unlikely,
see Muno et al. 2000).

RXTE/ASM data showed that the source entered a quiescent state in 
February 2001 (upper limit of 2 count/s corresponding to a luminosity 
of $< 3 \times 10^{36}$ ergs/s). A few months after (on 27 March 2001) 
KS 1731--260 was observed by Chandra: it was still in quiescence with a
0.5--10 keV bolometric luminosity of $\sim 10^{33}$ ergs/s (assuming a 
distance of 7 kpc) and an X-ray spectrum that is well described by a 
blackbody at a temperature of $\sim 0.3$ keV (Wijnands et al. 2001a). 
The same data were also analyzed by Rutledge et al. (2001); they fitted
the spectrum using a hydrogen atmosphere model, obtaining an effective 
temperature of 0.12 keV, an emission area radius of $\sim 10$ km and a 
bolometric luminosity of $\sim 2.7 \times 10^{33}$ ergs/s (assuming a 
distance of 8 kpc).
We present here the results of a 200 ks 
BeppoSAX observation of KS 1731--260 taken in March 2001, just after the
beginning of this quiescent state, in which the source was detected at a 
quiescence luminosity level of $(7 \pm 2) \times 10^{32}$ ergs/s.

\section{Observations and Data Analysis}

KS~1731--260 was observed with the BeppoSAX Narrow Field Instruments (NFI) 
from 2001 March 2 23:28 UT to March 5 19:07 UT for 200 ks. 
The effective exposure time was $\sim 91$ ks. 
The NFIs consist of four co-aligned instruments, namely a Low Energy 
Concentrator Spectrometer, LECS, a thin window position sensitive 
proportional counter with extended low 
energy response in the band 0.1--10~keV (Parmar et al. 1997,
due to UV contamination problems, the LECS was  
operated only at satellite night time, resulting in a reduced 
exposure time with respect to the other instruments);
two (originally three) Medium Energy Concentrator Spectrometers, MECS, 
operating in the 1.3--10~keV band (Boella et al. 1997);  
a High Pressure Gas Scintillation Proportional Counter, 
HPGSPC, sensitive in the energy range 7--60 keV (Manzo et al. 1997);
a Phoswich Detection System, PDS, sensitive in the energy range 13--200 keV 
(Frontera et al. 1997), consisting of four independent NaI(Tl)/CsI(Na) 
phoswich scintillation detectors. 

Images from the two MECS were summed and the events with PHA energies 
between 1.3 and 10 keV were considered. We searched for sources in the MECS 
field of view (FOV) around the nominal position of KS~1731--260 (Barret et al.
1998). We found a source centered at the position RA: 17h 34m 09.4s, 
DEC: $-26^\circ$ $06'$ $22''$, equinox 2000, and extracted all the photons 
from a $1.8'$ radius circle (corresponding to 64\% of the point spread 
function, PSF; Fiore, Guainazzi \& Grandi 1999). The size of the extraction 
region was determined to optimize the signal to noise ratio.
The accuracy and stability of the pointing during the BeppoSAX observation of 
KS 1731--26 has been checked and the satellite attitude has been verified 
within each orbit by telemetry reconstruction. In particular we have verified 
that the positions of the stars in the on board startrackers FOV 
remained constant during all the observation.
In principle, the error on the BeppoSAX/MECS position is $\sim 0.55'$
($1 \sigma$ confidence level); however, given the possibility of 
confusion with other unresolved sources (see below), we will consider
as a conservative estimation of the error on the source position
the radius, $1.8'$, of the extraction region.

The recently reported source position based on a Chandra observation performed 
on March 27 2001 and corrected for an offset with respect to the
2MASS (Wijnands et al. 2001b) is
RA: 17h 34m 13.47s,  DEC: $-26^\circ$ $05'$ $18.8''$, equinox 2000.
Therefore our derived position differs by $1.40'$ from the Chandra 
position, which is still within the $1.8'$ radius circle. 
Because of the position of the source close to the Galactic center,
the possibility of contamination from diffuse emission and other sources, 
which might fall within the $1.8'$ radius circle, must be considered. 
Indeed in the Chandra observation of KS~1731--260, another X-ray source 
(probably an X-ray active star) was detected at $0.5'$ from KS 1731--260
(Wijnands et al. 2001a, 2001b).  The MECS FOV, with the source we detected, 
is shown in Figure 1. The position of the Chandra sources are also indicated 
in the figure. Therefore the luminosity we will derive for KS 1731--260 
should be considered as an upper limit.  

With this caveat in mind, we derived the luminosity of the source in the 
following way.
Since standard background subtraction methods based on blank sky 
observations are not applicable, we extracted background counts from 
nearby source-free regions. 
The background was taken from a region within the 8 central arcmins
of the MECS FOV. We checked that the number of source photons estimated does
not vary, within 20\%, using background boxes of different sizes and 
locations. Of the 280 events detected within the $1.8'$ radius circle 
centered on the source position, 210 events ($\sim 75\%$ of the total) are 
expected to be background events. 
This corresponds to a $3.7 \sigma$ detection. 
From these data we extracted a light curve; no short term variability
is significantly detected during the BeppoSAX observation.
The count rate corrected for the PSF is $(1.20 \pm 0.33) \times 10^{-3}$
counts s$^{-1}$. Because the statistics in the MECS spectrum is not enough
to constrain the spectral model, we considered different spectral models, 
namely a blackbody (with temperatures ranging between 0.2 and 1 keV) and a 
power law (with photon indices ranging between 1.5 and 2.9).
In all cases the column density $N_{H}$ was fixed at  
$1.0 \times 10^{22}$ cm$^{-2}$, consistent with the value obtained from 
the ROSAT and Chandra observations (Barret et al. 1998; Wijnands et al. 2001a).  
Within 10\% all the models yield a flux of $\sim 1 \times 10^{-13}$ erg 
cm$^{-2}$ s$^{-1}$ in the $1.3-10$ keV band. 
For a blackbody of temperature 0.3 keV (in agreement with the blackbody
fitted to the Chandra data of KS~1731--260, Wijnands et al. 2001a), adopting 
a distance of 7 kpc (Muno et al. 2000), the corresponding bolometric 
luminosity is $(7 \pm 2) \times 10^{32}$ erg s$^{-1}$. This value is 
compatible with the $0.5-10$ keV bolometric luminosity of $\sim 10^{33}$
ergs/s deduced from the Chandra observation (Wijnands et al. 2001a).
As a cross check we also extracted the data in the MECS FOV using a circular 
region of $4'$ radius (within which $\sim 90\%$ of the source counts are 
expected) centered on the nominal position of KS 1731--260.  In this case
the background was measured from a region of the MECS image $20'$ away from 
the FOV center and divided by a factor of 0.85 to take into account the 
reduction of the effective area with the increase of the distance from 
the FOV center.
In this case the background contributes $\sim 80\%$ of the total source plus 
background spectrum.  The estimated bolometric (unabsorbed) source luminosity 
was $\sim 1.4 \times 10^{33}$ ergs/s using a blackbody model.
In the following we will adopt as a reasonable estimate for the 
source luminosity in quiescence $L_q \la 1 \times 10^{33}$ erg s$^{-1}$. 

At the beginning of August 2001, KS 1731--260 was observed, still in its
quiescence state (according to the ASM light curve), at 1.4 GHz with the 
Parkes radiotelescope to search for radio pulses.
Taking into account the possibility, suggested  by Muno et al. (2000),
that the NS belongs to an extremely narrow binary system, with  an
orbital period of $\sim 1$ hr,  we subdivided the data into segments  
of the duration of 35 and 4 min, respectively.   The analysis of longer
segments allows  to reach a  better sensitivity if the NS is 
in a wide orbit.  On the other hand, using shorter segments, the time of
arrival of the radio pulses is not heavily affected by the
strong Doppler effect caused by the orbital  motion in a narrow binary 
system, which could reduce the coherence of the signal.
We found that no radio signals with millisecond periods are present in the 
data above a flux limit of 0.21 mJy for the 35 min integration time, and 
above 0.60 mJy for the 4 min integration time.

\section{Discussion} 

The 90 ks BeppoSAX observation of KS 1731--260 carried out on 2001 
March 2--5 led to a detection of the source at a luminosity level of 
$\sim 10^{33}$ ergs/s. 
This is similar to the flux determined in a subsequent observation of the 
source carried out with Chandra on 2001 March 27 (Wijnands et al. 2001a).
These results testify that the quiescent X-ray luminosity and spectrum of 
KS 1731--260 are close to those determined for other NS SXTs. 
We therefore assume that the X-ray spectrum of KS 1731--260 in quiescence 
is similar to that of Aql~X-1 and Cen~X-4 (see e.g. Campana et al. 1998a), 
the best studied cases. 
The spectrum of these two sources could be well fit by a soft
thermal component (blackbody temperature of $\sim 0.2$~keV) plus
a power-law component with a photon index $\Gamma \sim 1.5$.
The blackbody component is usually interpreted as thermal emission from a pure
hydrogen NS atmosphere (e.g.\ Rutledge et al. 1999, 2000), while the 
power-law component is thought to be due to residual accretion or the 
interaction of a pulsar wind with matter released by the companion star
(see e.g.\ Campana \& Stella 2000 and references therein). 

In the following we discuss the properties of KS 1731--260 in relation 
to the different mechanisms that have been proposed to explain
the quiescent X-ray emission of NS SXTs. 
In summary, there exist three sources of energy which are expected to produce 
X-ray luminosity in quiescence: 
\begin{itemize}
\item[a)] residual accretion onto the NS surface at very low rates 
(e.g.\ Stella et al. 1994); 
\item[b)] rotational energy of the NS converted into radiation
through the emission from a rotating magnetic dipole, a fraction of
which can be emitted in X-rays (e.g.\ Possenti et al. 2001; 
Campana et al. 1998b, and references therein); 
\item[c)] thermal energy, stored into the NS during previous phases
of accretion, released during quiescence (e.g.\ Brown, Bildsten, \& 
Rutledge 1998; Colpi et al. 2001; Rutledge et al. 2001). 
\end{itemize}
We will compare the corresponding expectations with the measured upper limit
on the X-ray luminosity.
Note that the last process has to be always taken into account,
as the source was recently active for more than one year
at a luminosity of $\sim 10^{37}$ erg/s and frequently active, at comparable
luminosities, since its discovery in 1989.
Therefore we firstly derive some constraints resulting from the processes
a) and b), and then we combine these with the constraint derived from 
process c). 

Some constraints for the processes a) and b) can be derived if we
assume that the coherent pulsations detected during the type I X-ray
burst on 1996 July 14 (Smith et al. 1997) correspond to
the NS spin.  Indeed the high degree of coherence ($\nu /
\Delta \nu \ga 900$) is compatible with broadening 
induced by the finite duration of the oscillations ($\sim 2$ s).
This strongly supports the interpretation of the frequency of these
oscillations ($\nu = 523.92$ Hz) as the rotational frequency of the
NS. Note that the frequency separation between the kHz QPOs in KS 1731--260 
is $\sim 260$ Hz (Wijnands \& van der Klis 1997), which might be interpreted 
as the spin frequency of the NS (Strohmayer et al. 1996; Miller, 
Lamb, \& Psaltis 1998), while the burst frequency would be its first overtone 
in this case (see, however, Muno et al. 2000). 
For concreteness in the following we will assume the burst frequency as the 
NS spin frequency and will discuss differences when needed.

%
%
It has been demonstrated that a rotating magnetic dipole in vacuum
emits electromagnetic dipole radiation. Moreover a wind of relativistic 
particles associated with magnetospheric currents along the field lines
is expected to arise in a rotating NS (e.g.\ Goldreich \& Julian 1969). 
Both these processes, powered by the rotational energy of the NS, 
depend on the angle between the NS magnetic moment and the spin axis
and compensate in such a way that the total energy emitted is nearly 
independent of this angle (Bhattacharya \& van den Heuvel
1991). Thus the bolometric luminosity of a rotating NS in vacuum can be
calculated according to the Larmor's formula $L_{\rm bol}= (2/3c^3)\mu^2 
\omega^4$, where $c$ is the speed of light, $\mu = B_{\rm s} R_{\rm NS}^3$ 
(where $B_{\rm s}$ is the surface magnetic field at the magnetic equator and
$R_{\rm NS}$ is the NS radius), and $\omega$ is the angular frequency of the
NS. 
A small fraction of this luminosity is emitted in the radio band, probably
as the results of electron accelerations in ultra-strong electric potentials
(gaps). A still open question is the location of the accelerator (gap), 
in the outer magnetosphere, close to the light cylinder radius
($r_{\rm LC} = c P / 2 \pi$, i.e.\ the radius at which an object corotating 
with the NS attains the speed of light), as proposed for
the pulsar emission mechanism by Halpern \& Ruderman (1993), or
close to the magnetic cap, as originally proposed by Arons (1981) and 
Daugherty \& Harding (1982).
The observability radio emission can be strongly affected by the matter 
surrounding the NS. In particular free-free absorption can strongly reduce 
the radio flux especially at low radio frequencies (see e.g.\ Burderi \& King 
1994, and, more recently, D'Amico et al. 2001). 
However, independent of the observability of the radio pulsar,
this radiative regime certainly occurs when the space surrounding the NS
is free of matter up to $r_{\rm LC}$ and the pressure of this radiation
can overcome the pressure of the accretion flow, thus preventing further
accretion (see Ruderman, Shaham, \& Tavani 1989; Illarionov \& Sunyaev 1975).
Therefore, in the following, we will adopt the hypothesis 
that once the magnetospheric 
radius is outside the light cylinder radius (and thus the space 
surrounding the NS up to the light cylinder is free of matter) 
the NS emits radiation according to the Larmor's formula. 
In this case, irrespective of the amount of radiation emitted in the 
radio band and of its observability, we will say, for shortness, that the 
radio pulsar is active.
%
%

As mentioned above, there are two possibilities: the first possibility,
scenario a), is that the magnetospheric radius is inside the light cylinder 
radius 
and therefore the radio pulsar is off. In this case, for a non-zero
NS magnetic field, we should have some matter flow toward the NS in order 
to keep
the magnetospheric radius small enough. In this scenario we again have two
possibilities: a1) the magnetospheric radius is inside the co-rotation radius
(see below for a definition),
so accretion onto the NS surface is possible; a2) the magnetospheric radius is
outside the co-rotation radius (but still inside the light cylinder radius),
so the accretion onto the NS is centrifugally inhibited, but an accretion 
disk can still be present and emit X-rays.
The other possibility, scenario b), is that the magnetospheric radius is 
outside the light cylinder radius and 
therefore the radio pulsar is on. In this case X-ray emission can be 
produced by: b1) reprocessing of 
part of the bolometric luminosity of a rotating NS
(Larmor's radiation formula)
into X-rays in a shock
front; b2) the intrinsic X-ray emission of the radio pulsar.

Let us consider scenario a), i.e.\ a non-zero accretion rate during the 
quiescent phase, and in particular scenario a1).
If the NS has an intrinsic magnetic field, we can estimate an upper limit 
on the NS magnetic moment. 
Note that a similar argument was applied to an ASCA observation of the Rapid
Burster in quiescence by Asai et al. (1996). Also in that case it was 
concluded that a highly magnetized and rapidly rotating NS can be excluded.
The matter accreting onto the NS forms an accretion
disk whose inner radius is truncated at the magnetosphere by the
interaction of the accretion flow with the magnetic field of the NS.
In this case the magnetospheric radius $r_{\rm m}$ 
is a fraction $\phi \la 1$ (an expression for $\phi$ can be found in Burderi
et al. 1998\footnote{$\phi = 0.21 \alpha^{4/15} n_{0.615}^{8/27} m^{-142/945}
[(L_{37}/\epsilon)^{8/7} R_6^{8/7} \mu_{26}^{5/7}]^{4/135}$, see the text
for the definition of the symbols.}; 
for $L \sim 10^{33}$ ergs/s we get $\phi \sim 0.2$) of the Alfv\'en 
radius $R_{\rm A}$ defined as the radius at which the energy density of the 
(assumed dipolar) NS magnetic field equals the kinetic energy density of the 
spherically accreting (free falling) matter:
\begin{equation}
R_{\rm A} = 2.23 \times 10^6  R_6^{-2/7}  m^{1/7}
\mu_{26}^{4/7} \epsilon^{2/7} L_{37}^{-2/7} \;{\rm cm}
\end{equation}
(see, e.g., Hayakawa 1985),
where $R_6$ is the NS radius, $R_{\rm NS}$, in units of $10^6$ cm, $m$ is the 
NS mass in solar masses, $\mu_{26}$ is the NS magnetic moment in units of
$10^{26}$ G cm$^3$, 
$\epsilon$ is the ratio between the specific luminosity
and the specific binding energy ($L = \epsilon \times G M \dot{M}/R_{\rm NS}$,
$G$ is the gravitational constant, $M$ is the NS mass, $\dot{M}$ is
the accretion rate), and $L_{37}$ is the accretion luminosity in units of
$10^{37}$ erg/s, respectively.  
Actually, accretion onto a spinning, magnetized NS
is centrifugally inhibited once the magnetospheric radius is outside
the corotation radius, the radius at which the Keplerian frequency 
of the orbiting matter is equal to the NS spin frequency:
\begin{equation}
r_{\rm co} = 1.5 \times 10^6 P_{-3}^{2/3} m^{1/3} \;{\rm cm}
\end{equation}
where $P_{-3}$ is the spin period in ms.
The condition to allow accretion then reads $r_{\rm m}/r_{\rm co}
\le 1$. 
This requires:
\begin{equation}
\mu_{26} \le 0.5 \phi^{-7/4} R_6^{1/2} \epsilon^{-1/2} L_{37}^{1/2} 
	     m^{1/3} P_{-3}^{7/6}.  
\end{equation}
Adopting our upper limit, $L_{37} = 10^{-4}$, for the quiescent luminosity
we obtain $\mu_{26} \le 0.28 m^{1/3}$ for $P_{-3} = 1.91$
($\mu_{26} \le 0.63 m^{1/3}$ for $P_{-3} = 3.82$), where
we have assumed $\epsilon = R_6 = 1$ and $\phi \simeq 0.2$.

If the magnetospheric radius is outside the corotation radius and still 
inside the light cylinder radius (scenario a2)
the accretion is centrifugally inhibited, but the disk can still 
emit X-rays.
This will increase the upper limit on the magnetic field derived above
because in the propeller regime the system will be under-luminous, by a 
factor $\epsilon_{\rm prop} = R_{\rm NS}/(2 r_{\rm disk})$, for a given 
accretion rate. 
In this case, the maximum accretion rate for a given luminosity occurs
when the efficiency factor $\epsilon$ is minimum.  This occurs when the 
inner disk radius is the furthest from the NS surface without allowing the 
switch-on of the radio pulsar, i.e. for $r_{\rm disk} \la r_{\rm LC}$.
Assuming $r_{\rm m} = r_{\rm LC}$ and taking as upper limit on $\dot M$ the
condition $L_q = L_{\rm disk} (r_{\rm LC}) = \epsilon_{\rm prop} 
G M \dot{M}/R_{\rm NS}$, with $\epsilon_{\rm prop} = 
R_{\rm NS}/(2 r_{\rm LC})$, we can calculate the upper limit on the magnetic 
moment in this case:
\begin{equation}
\mu_{26} \le 11.7 \phi^{-7/4}  L_{37}^{1/2}  P_{-3}^{9/4} m^{-1/4},
\end{equation}
where we have assumed $R_6 = 1$.
This gives $\mu_{26} \le 8.4 m^{-1/4}$ 
for $P_{-3} = 1.91$
(and $\mu_{26} \le 39.8 m^{-1/4}$
for $P_{-3} = 3.82$), where
we have assumed $L_{37} = 10^{-4}$ for the quiescent luminosity, and
$\phi \simeq 0.2$ (in fact, as $\phi \propto \epsilon_{\rm prop}^{-32/945}$,
$\phi = 0.18 \simeq 0.2$ is also appropriate in the propeller case). 

Let us consider now the scenario b). If there is no accretion in quiescence
or the magnetospheric radius falls outside the light cylinder radius, we 
expect that the radio pulsar switches on.
The pulsar radiation pressure may overcome the pressure of the 
accretion disk, thus determining the destruction of the disk and the 
ejection of matter from the system (see Burderi et al. 2001).  
However there is the possibility that a fraction $\eta$ of this power can 
be converted into X-rays in a shock front between the emerging radiation 
and the circumstellar matter (scenario b1).
Typical values for $\eta$ are in the range $\sim 0.01-0.1$ (e.g.\
Campana et al. 1998b). In this case $\eta L_{\rm rad} \la L_q$ or 
\begin{equation}
\mu_{26} \le 5.1 L_{37}^{1/2} P_{-3}^2 \eta^{-1/2}.
\end{equation}
In our case, assuming the minimum value for the efficiency in the conversion
of the rotational energy into X-rays,
$\eta = 0.01$, which implies the highest value for the magnetic moment, 
we get $\mu_{26} \le 1.9$ for $P_{-3} = 1.91$
(and $\mu_{26} \le 7.5$ for $P_{-3} = 3.82$).
Let us consider the possibility that part of the spin-down energy 
loss is directly emitted in X-rays (scenario b2). Indeed about 40 out of 
$\sim 1400$ known radio pulsars have been detected in the X-ray range so far. 
In this case a correlation has been observed between the X-ray and spin-down 
luminosities (e.g.\ Becker \& Trumper 1997; Possenti et al. 2001, and
references therein).  Considering the empirical relation, derived by
Possenti et al. (2001) analyzing a sample of 37 pulsars, between the
2--10 keV luminosity, $L_{37}$, and the rate of spin-down energy loss, 
$L_{\rm rad}$, 
\begin{equation}
L_{37} = 2.51 \times 10^{-52} L_{\rm rad}^{1.31},
\end{equation}
we can again calculate an upper limit on the magnetic moment in the 
case of KS 1731--260,
which gives $\mu_{26} \le 8.7$ for $P_{-3} = 1.91$ (and $\mu_{26} \le 34.6$
for $P_{-3} = 3.82$).

%
%
In the last two cases, the X-ray emission is expected to be non-thermal 
(a power-law spectrum could be appropriate). Interestingly Wijnands
et al. (2001) found that the Chandra spectra of KS 1731--260 in quiescence
could be equally well fitted by a blackbody or a power law plus a blackbody
(although the power law is not statistically required).
In this second case one might think that the thermal component arises
from the NS surface and the non-thermal power law is associated with
the shock emission discussed above. Indeed, according to the results of
the spectral deconvolution of Wijnands et al. (2001), the fraction of the total
luminosity emitted in the power law is $\sim 15\%$, small as compared to the 
total luminosity. However, because of the low statistics, the spectral
deconvolution is not secure and, in the case of our BeppoSAX observation,
no statistically significative spectral analysis is possible.
For this reason, in deriving our upper limits on the magnetic moment strength
in these scenarios, we have adopted the very conservative assumption that
the whole flux detected by BeppoSAX can be ascribed to the non-thermal emission.
%
%

An upper limit on the NS magnetic field can also be derived 
from the upper limit on the radio pulsed emission from KS 1731--260.
Radio observations of KS 1731--260, taken at the Parkes radiotelescope
while the system was in quiescence,
showed that the upper limit on the flux of pulsed radio emission is 
0.21 mJy using 35 min integration time and 0.60 mJy using 4 min integration
time.  Equating the rotational energy lost by a magnetized NS  with its
magnetic dipole radiation, we obtain:
\begin{equation}
P \dot{P} = 9.75 \times 10^{-24} \mu_{26}^2 I_{45}^{-1}
\label{eq:ppdot}
\end{equation}
where $P$ and $\dot{P}$ are respectively the period of the radio 
pulsar and its derivate 
and $I_{45}$ is the NS momentum of inertia in units of $10^{45}$ g cm$^2$.
Proszynski \& Przyibician (1985), using a sample of 275 pulsars, found 
an empirical relation between the observed 400 MHz luminosity, $L_{400}$, 
and the pulsar parameters:
\begin{equation}
\log(L_{400})=\frac{1}{3} \log\left(\frac{\dot{P}_{-15}}{P^3}\right) + A
\label{eq:L400}
\end{equation}
where $L_{400}$ is expressed in  mJy kpc$^2$, $\dot{P}_{-15}$ is in
units of $10^{-15}$ s s$^{-1}$ and $A  = 1.1$.  
Kulkarni,  Narayan  \&  Romani (1990) found that the  same
relation could easily fit a sample of 11 recycled binary pulsars.
For KS 1731--260, assuming a distance of 7.0 kpc and a spectral index
$\alpha = 1.7$, typical of millisecond pulsars, we obtain for the 
400 MHz luminosity an upper limit of 88.47 mJy kpc$^2$ in the case of 35 min
integration, and of 250.25  mJy kpc$^2$ in the case of 4 min integration.
Using  these values    and  combining  equations   \ref{eq:ppdot}  and
\ref{eq:L400}, we can again calculate  an upper limit  to  the
magnetic field of KS 1731--260. If the NS  is in a  wide orbit, i.e. if
the phase shifts introduced  by the doppler effect are negligible
over  an integration time  of 35 min,  we obtain  an  upper limit of 
$\mu_{26} \le 0.68$ for $P_{-3} = 1.91$ (and $\mu_{26} \le 2.7$
for $P_{-3} = 3.82$), assuming
$I_{45}=R_6=1$. If, instead, the NS in KS 1731--260 belongs to a narrow
binary  system and then the 4 min observations are  more appropriate, we
obtain $\mu_{26} \le 3.1$ for $P_{-3} = 1.91$ 
(and $\mu_{26} \le 12.6$ for $P_{-3} = 3.82$).
Note however that free-free absorption from matter surrounding the
system (expected if the mass loss from the companion proceeds at some
level during the quiescent phase, see Burderi et al. 2001
for a detailed discussion of this possibility in transient systems)
can efficiently hamper the detection of the pulsed radio emission
from an active radio pulsar. 

We will consider now the process c), i.e.\ the thermal emission from the
NS. Deep crustal heating was first developed by Haensel \& Zdunick
(1990).  It was pointed out by Brown et al. (1998) that this scenario
was relevant to accreting NSs in transient systems, and may contribute to 
the emission at low luminosities.  This scenario applies if the system
has time to reach a steady state, in which the heat deposited during short 
and frequent outbursts in the NS is equal to the luminosity radiated in 
quiescence.
Colpi et al. (2001) have explored the thermal evolution of a NS 
undergoing episodes of accretion, lasting for a time $t_{\rm out}$, 
separated by periods of quiescence, lasting $t_{\rm rec}$ in the hypothesis
that the thermal luminosity in quiescence is dominated by emission from 
the hot NS core, with which the crust and atmosphere of the NS 
are in thermal equilibrium.  
Adopting an outburst luminosity of $10^{37}$ ergs/s and a quiescent
luminosity of $10^{33}$ ergs/s, this model predicts a ratio of the recurrence
time $t_{\rm rec}$ to the outburst time $t_{\rm out}$ of $\sim 70$ for a 1.4
$M_\odot$ NS and less than 4 for a NS more massive than 1.7 $M_\odot$.
A $t_{\rm rec}/t_{\rm out} \sim 70$ would imply that, on average, the total 
time the source should spend in outburst in 13 yrs (i.e.\ since its discovery) 
should be about 2 months. 
This is in conflict with time history of this source and 
the fact that the last outburst episode observed by the ASM lasted much more 
than one year. As already pointed out by Wijnands et al. (2001a), we are 
therefore left with two possibilities: 
i) we are witnessing a peculiar period of hyperactivity of the source, which 
will be followed by a much longer (up to a thousand years);
ii) the NS in this system is quite massive, in agreement with the expectation 
of the standard conservative recycling scenario (e.g.\ Bhattacharya \& 
van den Heuvel 1991) that the NS has accreted a significant amount of mass 
and has been spun-up to a millisecond spin period.  

However, Rutledge et al. (2001) found that, since KS 1731-260 has recently 
experienced an extremely long outburst, the thermal 
luminosity in quiescence could be dominated by emission from the heated NS 
crust out of thermal equilibrium with the core.  Indeed the prediction
of the quiescent luminosity from the cooling crust model agrees (within a 
factor of a few) with the observed bolometric luminosity in quiescence. 
In this case the model above cannot be applied to the system, given that 
the core might be much cooler than indicated by the thermal luminosity in 
quiescence.  This implies that the presence of a massive NS is not required
to explain the behavior of this system, and that the estimated long 
outburst recurrence timescale mentioned above should be considered as a 
lower limit.
Rutledge et al. (2001) also calculated the expected thermal evolution 
of the crust, which should cool in 1--30 years, and therefore the time
evolution of the quiescent luminosity for different crustal conductivities 
and cooling processes, showing that a monitoring of this source can provide 
valuable information on the crust and core microphysics. 

A further constraint on the magnetic moment can be derived considering 
that the NS spin period must be longer than the equilibrium spin period,
namely the Keplerian period at the magnetospheric radius:
\begin{equation}
P_{\rm eq} = 1.81 \times 10^{-3} \phi^{3/2} R_6^{-3/7} 
\mu_{26}^{6/7} \epsilon^{3/7} L_{37}^{-3/7} m^{-2/7} \; {\rm s.}
\end{equation}
Adopting $\phi = 0.2$, and an outburst luminosity of $L_{37} = 1$ 
(i.e.\ the averaged luminosity observed with ASM and previous missions when 
the source was detected) 
we obtain $\mu_{26} \le 16.4 m^{1/3}$ for $P_{-3} = 1.91$ 
(and $\mu_{26} \le 36.8 m^{1/3}$ for $P_{-3} = 3.82$).
As KS~1731--260 is a transient source it is possible that, during quiescence,
the NS enters a propeller phase during which an efficient torque
could spin down the NS far from the equilibrium period defined above.
In this case the magnetic moment is significantly {\it less} than
the value derived in this paragraph. On the other hand, if the propeller
phases are infrequent and/or the spin down torque is unefficient, the 
NS is expected to be in spin equilibrium and $\mu_{26} = 16.4 m^{1/3}$
($\mu_{26} = 36.8 m^{1/3}$).
 
From the discussion above we conclude that, in any case, the NS magnetic 
field is most probably less than $\sim 10^{9}$ Gauss, 
and less than $\sim 4 \times 10^{9}$ Gauss in the worst case.

Let us finally discuss the consequences of the spin equilibrium constraint
when applied to the proposed scenarios in quiescence.
In fact, if we consider the case that the NS in KS~1731--260 is quite 
massive, the amount of mass accreted, $\ga 0.3$ $M_\odot$, is
sufficient, in principle, to spin up the NS below 1 ms for soft or 
moderately stiff equations of state for the NS matter and below 1.5 ms 
even for the stiffest equations of state (see Burderi et al. 1999).
Therefore:
\begin{itemize}
\item[a1)]
The NS in quiescence accretes matter onto its surface at very low rates.
In this case $\mu_{26} \le 0.28 m^{1/3}$ ($\mu_{26} \le 0.63 m^{1/3}$).
This, compared with the spin equilibrium constraint, i.e.\ 
$\mu_{26} \le 16.4 m^{1/3}$ ($\mu_{26} \le 36.8 m^{1/3}$), implies that 
the NS is far from spinning at equilibrium, {\it i.e.} propeller phases 
must be frequent and the spin down torque very effective, although this 
quiescent phase does not correspond to a propeller.
\item[a2)]
The NS in quiescence is in a propeller phase. 
In this case $\mu_{26} \le 8.4 m^{-1/4}$ ($\mu_{26} \le 39.8 m^{-1/4}$).
This, compared with the spin equilibrium constraint, $\mu_{26} \le 16.4 
m^{1/3}$ ($\mu_{26} \le 36.8 m^{1/3}$), indicates that the NS is compatible 
with spinning close to the equilibrium. In this case propeller phases must be 
rare (although the present quiescent phase does correspond to a propeller) 
and/or the spin down torque not very effective.
\item[b1)]
The NS in quiescence is not accreting matter and the radio pulsar is active;
a fraction of the power emitted by the rotating NS is 
converted into X-rays in a shock front between the emerging radiation 
and the circumstellar matter.
In this case $\mu_{26} \le 1.9$ ($\mu_{26} \le 7.5$).
This, compared with the spin equilibrium constraint, implies that the 
NS is far from spinning at equilibrium, {\it i.e.} again propeller phases 
must be frequent and the spin down torque very effective, although this 
quiescent phase does not correspond to a propeller.
\item[b2)]
The NS in quiescence is not accreting matter and the radio pulsar is active;
in this case the fraction of power emitted by the rotating NS that is 
converted into X-rays in a shock front between the emerging radiation 
and the circumstellar matter is negligible, and the X-ray emission
is the intrinsic emission from the rotating NS.
In this case $\mu_{26} \le 8.7$ ($\mu_{26} \le 34.6$).
This implies that the NS is compatible with spinning close
to the equilibrium, {\it i.e.} propeller phases must be rare (indeed
the present quiescent phase does not correspond to a propeller) and/or 
the spin down torque not very effective.
\end{itemize}

In conclusion, scenarios a2) and b2) seems to be the most reasonable,
and both indicate that the NS is compatible with spinning close to the 
equilibrium period.

\acknowledgments
The authors would like to thank R.E. Rutledge and the anonymous referee
for enlightening discussions during the preparation of this work. 
This work was partially supported by a grant from the Italian Ministry of 
University and Research (Cofin-99-02-02).

\clearpage

\begin{figure}
\plotone{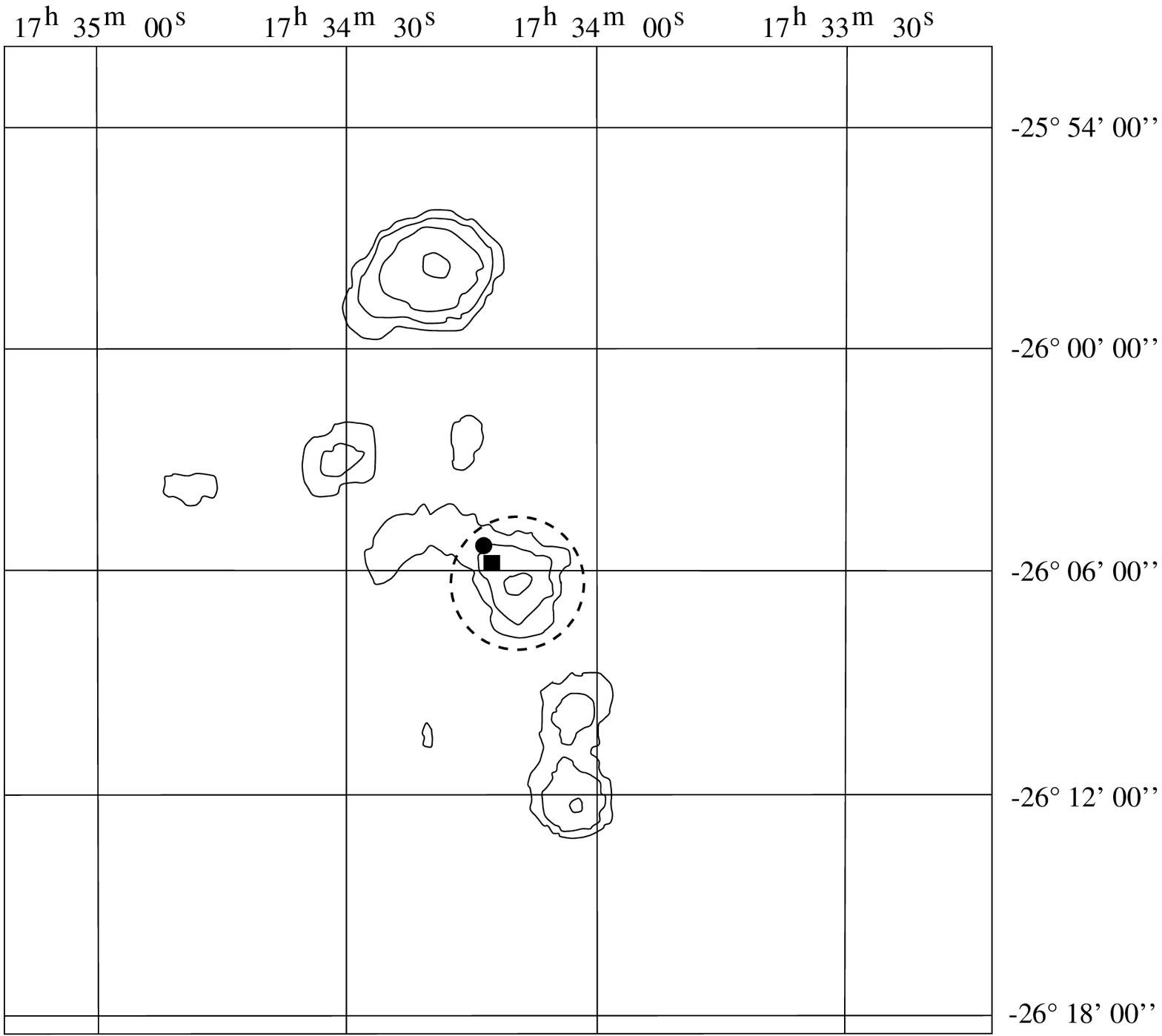}
\caption{\label{fig:fig1} 
Plot of the MECS field of view. The source we have detected close to the
nominal position of KS 1731--260 is shown at the center. There the dashed line
represents the $1.8'$ circular region from which the data have been extracted.
The positions of the sources identified in the Chandra observation are also
shown, i.e. KS 1731--260 (filled circle) and the 2MASS star (filled square).}
\end{figure}

\end{document}